\documentclass[aps,prl,twocolumn,showpacs,superscriptaddress,citeautoscript]{revtex4-1} 
\usepackage{graphicx}
\usepackage{amsmath,amssymb,amsfonts}
\usepackage{textcomp}
\usepackage{hyperref}
\usepackage{gensymb}
\usepackage{verbatim}
\usepackage{amsmath}
\hypersetup{breaklinks=true,colorlinks=true,urlcolor=black}
\usepackage{color}

\newcommand{\CeTiGe} {CeTiGe$_3$}
\newcommand{\MPa} {MP1}
\newcommand{\MPb} {MP2}
\newcommand{\MPc} {MP1\textquotesingle{}}

\begin{document}

	\title{Quantum tricritical point in the temperature-pressure-magnetic field phase diagram of CeTiGe$_3$}
	\author{Udhara~S. \surname{Kaluarachchi}}  
	\affiliation{The Ames Laboratory, US Department of Energy, Iowa State University, Ames, Iowa 50011, USA}
	\affiliation{Department of Physics and Astronomy, Iowa State University, Ames, Iowa 50011, U.S.A.}
	\author{Valentin \surname{Taufour}$^*$}
	\affiliation{The Ames Laboratory, US Department of Energy, Iowa State University, Ames, Iowa 50011, USA}
	\author{Sergey~L. \surname{Bud'ko}}  
	\affiliation{The Ames Laboratory, US Department of Energy, Iowa State University, Ames, Iowa 50011, USA}
	\affiliation{Department of Physics and Astronomy, Iowa State University, Ames, Iowa 50011, U.S.A.}
	\author{Paul~C. \surname{Canfield}}
	\affiliation{The Ames Laboratory, US Department of Energy, Iowa State University, Ames, Iowa 50011, USA}
	\affiliation{Department of Physics and Astronomy, Iowa State University, Ames, Iowa 50011, U.S.A.}

	\date{\today}

	\begin{abstract}
	We report the temperature-pressure-magnetic field phase diagram of the ferromagnetic  Kondo-lattice CeTiGe$_3$ determined by means of electrical resistivity measurements. Measurements up to $\sim$\,5.8\,GPa reveal a rich phase diagram with multiple phase transitions. At ambient pressure, CeTiGe$_3$  orders ferromagnetically at $T_\text{C}$\,=\,14\,K. Application of pressure suppresses $T_\text{C}$, but a pressure induced ferromagnetic quantum criticality is avoided by the appearance of two new successive transitions for $p$\,$>$\,4.1\,GPa that are probably antiferromagnetic in nature. These two transitions are suppressed under pressure, with the lower temperature phase being fully suppressed above 5.3\,GPa.  The critical pressures for the  presumed quantum phase transitions are $p_1$\,$\cong$\,4.1\,GPa and $p_2$\,$\cong$\,5.3\,GPa. Above 4.1\,GPa, application of magnetic field shows a tricritical point evolving into a wing structure phase with a quantum tricritical point at 2.8\,T at 5.4\,GPa, where the first order antiferromagnetic-ferromagnetic transition changes into the second order antiferromagnetic-ferromagnetic transition.   

	\end{abstract}
\maketitle

	\section{Introduction}

	Quantum phase transitions (QPT) in metallic ferromagnets have been studied for many years and remain a subject of great current interest\,\cite{Brando2016RMP}. The paramagnetic (PM) to ferromagnetic (FM) transition can be suppressed with nonthermal control parameters such as pressure, chemical composition or external field often leading to a $T$\,=\,0\,K, QPT. However, according to the current theoretical models, when suppressing the FM phase with a clean parameter such as pressure,  a continuous PM to FM transition is not possible. Instead, the transition becomes of the first order or a modulated magnetic phase can appear. The possibility of a first-order transition or the appearance of modulated magnetic phases was first discussed in Ref.\,\onlinecite{Belitz1997PRB} and \onlinecite{Belitz1999PRL}. In the case of the transition becoming of the first order, a wing structure was predicted in Ref.\,\onlinecite{Belitz2005PRL} and observed in UGe$_2$\,\cite{Taufour2010PRL} and ZrZn$_2$\,\cite{Kabeya2012JPSJ}. The case of the appearance of a modulated magnetic phase is more complex\cite{Belitz1999PRL,Belitz1997PRB,Chubukov2004PRL,Conduit2009PRL,Karahasanovic2012PRB,Thomson2013PRB,Pedder2013PRB,Taufour2016PRL} and an experimental examples were found in LaCrGe3\,\cite{Taufour2016PRL} and CeRuPO\,\cite{Kotegawa2013JPSJ}. Observation of both tricritical wings and modulated magnetic phase in LaCrGe$_3$ is a good example of a complex phase diagram and provides a new example of the richness of the phase diagram of metallic quantum ferromagnets\,\cite{Kaluarachchi2017NatComm}. Recently, Belitz and Kirkpatrick proposed that such complex phase diagram is due to quantum fluctuation effects\,\cite{Belitz2017arXiv}.


 	Cerium based compounds have attracted attention due to interesting ground states, such as heavy-fermion, unconventional superconductor\,\cite{Grosche1996,Mathur1998Nature}, Kondo insulator\,\cite{Hundley1990PRB}, magnetic ordering\,\cite{Iglesias1997PRB,Evans1991}, etc. Whereas many Ce-based compounds manifest an antiferromagnetic (AFM) ground state, only  few systems are known with FM order and pronounced Kondo effects. CeRuPO\,\cite{Kotegawa2013JPSJ}, CeAgSb$_2$\,\cite{Myers1999JMMM,Sidorov2003PRB},  CeNiSb$_3$\,\cite{Sidorov2005PRB}, CePd$_2$Ge$_3$\,\cite{Burghardt1997} and Ce$_2$Ni$_5$C$_3$\,\cite{Yamada2010} are some examples of the Ce-based ferromagnets, which show complex phase diagrams under the application of pressure. Interestingly, the FM transition in these materials is suppressed with the pressure and new magnetic (most probably AFM) phases appear before the Curie temperature reaches 0\,K but no wing structure in the $T$-$H$-$p$ phase diagrams has been observed so far. According to the recent theoretical work by Belitz\,{\em et al.}\,\cite{Belitz2017arXiv},  it is possible to have unobservable tricritical wings inside the AFM dome. In most of these cases, lack of in-field measurements under pressure prevents from constructing the temperature-pressure-field phase diagram and getting a better understanding of the system. Therefore, it is interesting to further investigate the temperature-pressure-field effect on a Ce-based ferromagnetic system. To address this, we present measurements of electrical resistivity under pressure up to $\sim$\,5.8\,GPa and magnetic field up to 9\,T on ferromagnetic \CeTiGe{}.

	\CeTiGe{} is one of the relatively rare examples of a ferromagnetic Kondo lattice ($\gamma$=75\,mJ\,mol$^{−1}$\,K$^2$\,\cite{Inamdar2014}); it orders  with a Curie temperature, $T_\text{C}$\,=\,14\,K\,\cite{Manfrinetti2005}. It crystallizes in the hexagonal perovskite BaNiO$_3$\,-\,type structure ($P6_3/mmc$)\,\cite{Manfrinetti2005}. Magnetization measurements show highly anisotropic behavior with $c$-axis being the easy axis of magnetization\,\cite{Inamdar2014}. A Curie-Weiss fit to the susceptibility data yields an effective moment of 2.5\,$\mu_\text{B}$, consistent with the reported values\,\cite{Inamdar2014} and nearly equal to the value for free-ions trivalent Ce (2.54\,$\mu_\text{B}$). The reported saturation moment at 2\,K from the magnetization data (1.72\,$\mu_\text{B}$/Ce) along the $c$-axis\,\cite{Inamdar2014} is comparable with the value obtained from the neutron diffraction study(1.5\,$\mu_\text{B}$/Ce)\,\cite{Kittler2013PRB}. Substitution of titanium  by vanadium   (CeTi$_{1-x}$V$_x$Ge$_3$) causes a suppression of the Curie temperature down to 3\,K at $x$\,=\,0.3 and suggests a possible quantum critical point or phase transition near $x\approx$\,\,0.35  \,\cite{Kittler2013PRB}. In contrast to the effect of substitution, a very small, initial positive pressure derivative of $T_\text{C}$ (d$T_\text{C}$/d$p$\,$\approx$\,0.3\,K\,GPa$^{-1}$ up to 1\,GPa) suggests that \CeTiGe{} is located near the maximum of the magnetic ordering temperature in the Doniach model\,\cite{Kittler2013PRB}. However, all substitution and pressure measurements have been done on the polycrystalline material and only to modest pressure, $p$\,$<$\,1\,GPa. To get a better understanding of $T$-$p$-$H$ phase diagram, possible FM instability and QCP it is important to perform high pressure studies on single crystalline samples of \CeTiGe{} over a wide pressure  range.

	\section{Experimental Methods}

		Single crystals of  \CeTiGe{} were grown using a high temperature solution growth technique\,\cite{CANFIELD1992,Canfield2001}. A mixture of elemental Ce, Ti and Ge was placed in a 2 mL fritted alumina crucible\,\cite{Canfield2016} with a molar ratio of Ce:Ti:Ge\,=\,4:1:19\,\cite{Inamdar2014} and sealed in a silica ampule under a partial pressure of high purity argon gas. The sealed ampule was heated to 1200\,\celsius\, over 10 hours and held there for 5 hours. It was cooled to 900\celsius\, over 120 hours and excess liquid was decanted using a centrifuge. A good quality sample (based on the residual resistivity ratio)  for the pressure study was selected after ambient pressure characterization by the magnetization and resistivity measurements. Temperature and field dependent resistance measurements were carried out using a Quantum Design (QD) Physical Property Measurement System (PPMS) from 1.8\,K to 300\,K. The $ac$-resistivity ($f$\,=\,17\,Hz) was measured by the standard four-probe method with the 1\,mA current in the $ab$ plane. Four Au wires with  diameters of 12.5\,$\mu$m were spot welded to the sample.  A magnetic field, up to 9\,T, was applied along the $c$-axis, which corresponds to the magnetization easy axis\,\cite{Inamdar2014}. A modified Bridgman cell\,\cite{Colombier2007RSI} was used to generate pressure for the resistivity measurement. A 1:1 mixture of $n$-pentane:iso-pentane was used as a pressure medium. The solidification of this medium occurs around $\sim$6-7\,GPa at room temperature\,\cite{Tateiwa2009RSI,Piermarini1973JAP,Klotz2009,Kim2011PRB,Torikachvili2015RSI}. The pressure at low temperature was determined by the superconducting transition temperature of Pb\,\cite{Bireckoven1988JPESI}.

	\section{Results and Discussion}
		
		\begin{figure}[htb!]
			\begin{center}
				\includegraphics[width=8.5cm]{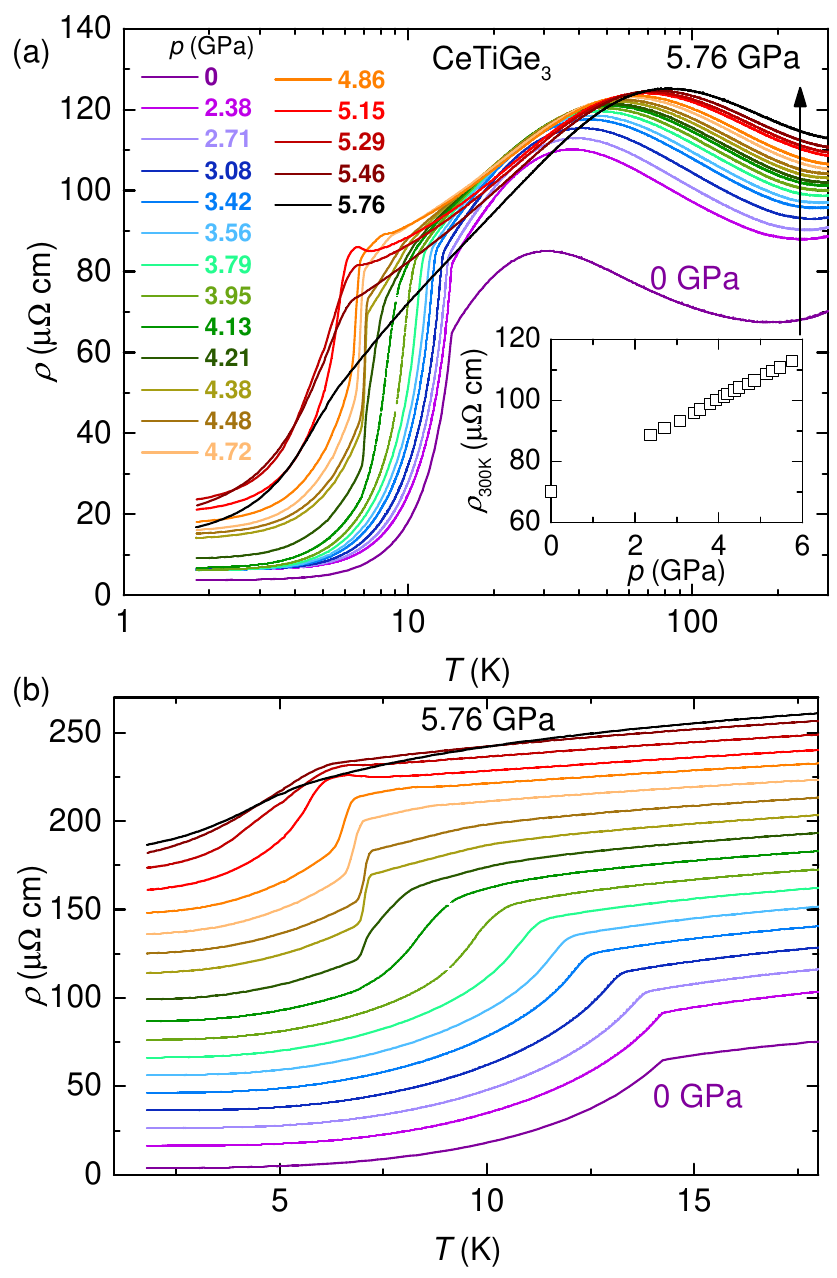}
			\end{center}
			\caption{\label{Rho_T}(Color online) (a) Temperature dependence of the in-plane resistivity, $\rho$($T$), of  a \CeTiGe{} single crystal under various pressures, $p$, up to 5.76\,GPa on a semi-log plot. The resistivity at 300\,K linearly increase with the pressure at a rate of 7.4\,$\mu\Omega$\,cm\,GPa$^{-1}$ from 0 to 5.76\,GPa as shown in the inset. (b) Low temperature resistivity at various pressures. Data are offset by increments of 10\,$\mu\Omega$\,cm for clarity.} 
		\end{figure}
		
		The temperature dependencies of the in-plane resistivity of single crystalline \CeTiGe{}  under various pressures up to 5.76\,GPa are shown in Fig.\,\ref{Rho_T}\,(a). At ambient pressure, the resistivity exhibits typical Kondo-lattice behavior with a broad minimum  $\sim$\,190\,K followed by a maximum at $T_\text{max}$\,=\,31\,K. The $T_\text{max}$ is assumed to be related to the Kondo interaction with a changing population of crystal electric field levels\,\cite{Inamdar2014,Cornut1972PRB,Hanzawa1985,Taufour2013PRB}. The FM transition manifests itself in the resistivity data as a sharp drop  at  $T_\text{C}$\,=\,14.2\,K. Similar values of $T_\text{C}$ have been reported from polycrystalline and single crystalline samples\,\cite{Manfrinetti2005,Kittler2013PRB,Inamdar2014}.  The residual resistivity ratio (RRR) is 19, a value that suggests a rather good quality of the sample. Upon application of pressure the resistivity at room temperature increases linearly with a rate of 7.4\,$\mu\Omega$\,cm\,GPa$^{-1}$ over the  whole pressure range (see inset of Fig.\,\ref{Rho_T}\,(a)), both the local  maximum  and local minimum in the resistivity broaden and move to higher temperatures with increasing pressure. The evolution of the low temperature resistivity is shown in Fig.\,\ref{Rho_T}\,(b); data are offset by increments of 10\,$\mu\Omega$\,cm for clarity.
		
		\begin{figure}[htb!]
			\begin{center}
				\includegraphics[width=8.5cm]{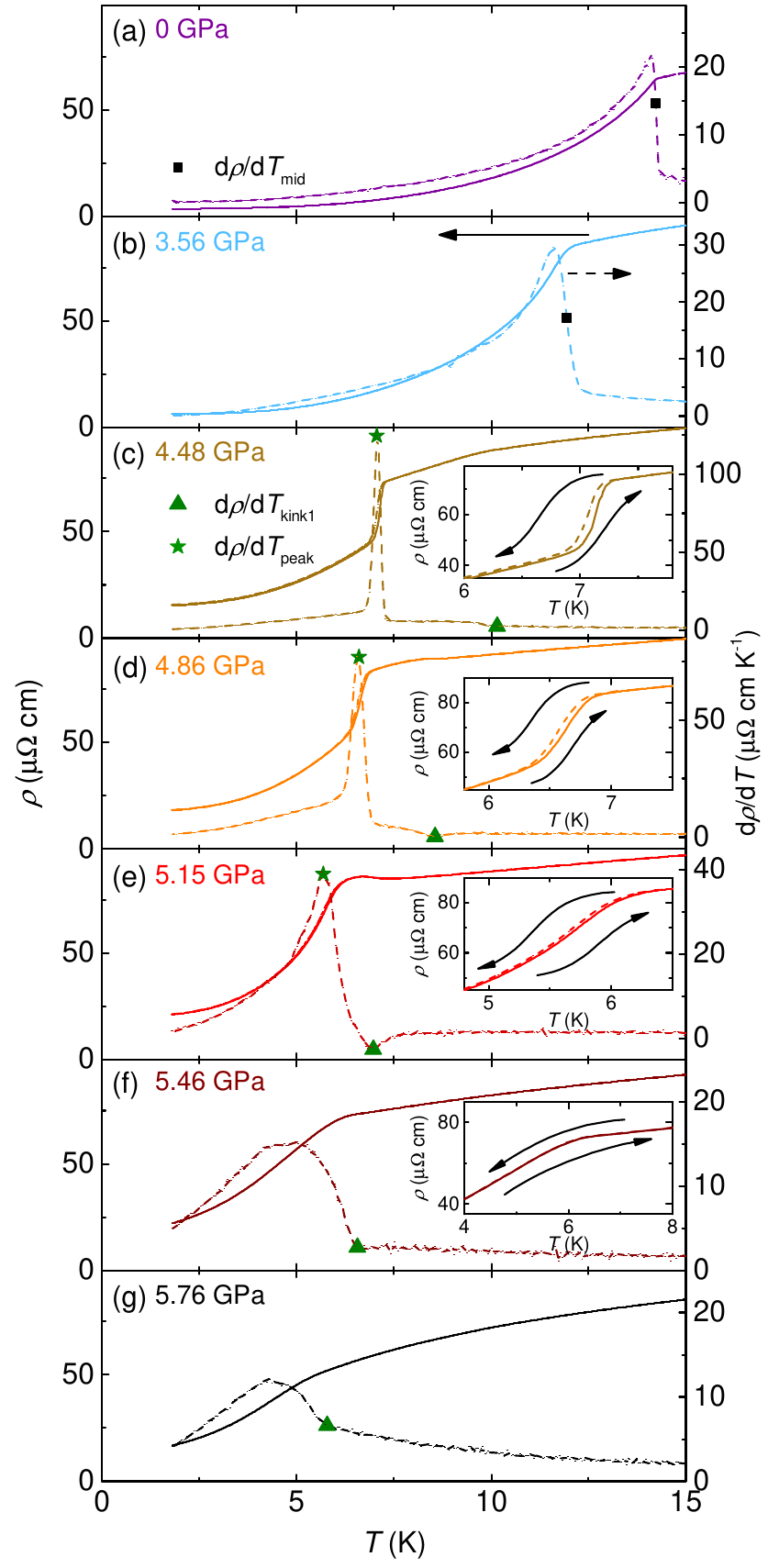}
			\end{center}
			\caption{\label{Rho_T20K}(Color online) Low temperature, in-plane resistivity (left axis) and its corresponding temperature derivative (right axis) of \CeTiGe{} for several representative pressure regions (a)-(b) $p$\,$<$\,4.1\,GPa, (c)-(e) 4.1\,GPa\,$<$\,$p$\,$<$\,5.3\,GPa and (f)-(g) 5.3\,GPa\,$<$\,$p$.  The solid symbols mark the characteristic temperatures that are associated with phase transitions: black square-PM to FM, green triangle- PM to \MPa{}/\MPc{}  and green star- \MPa{} to \MPb{}. The insets of (c)-(e) show the observed hysteretic behavior at their representative pressures. However, no hysteretic behavior is observed above 5.3\,GPa as shown in inset of (f).} 
		\end{figure}

\begin{figure}[htb!]
			\begin{center}
				\includegraphics[width=8.5cm]{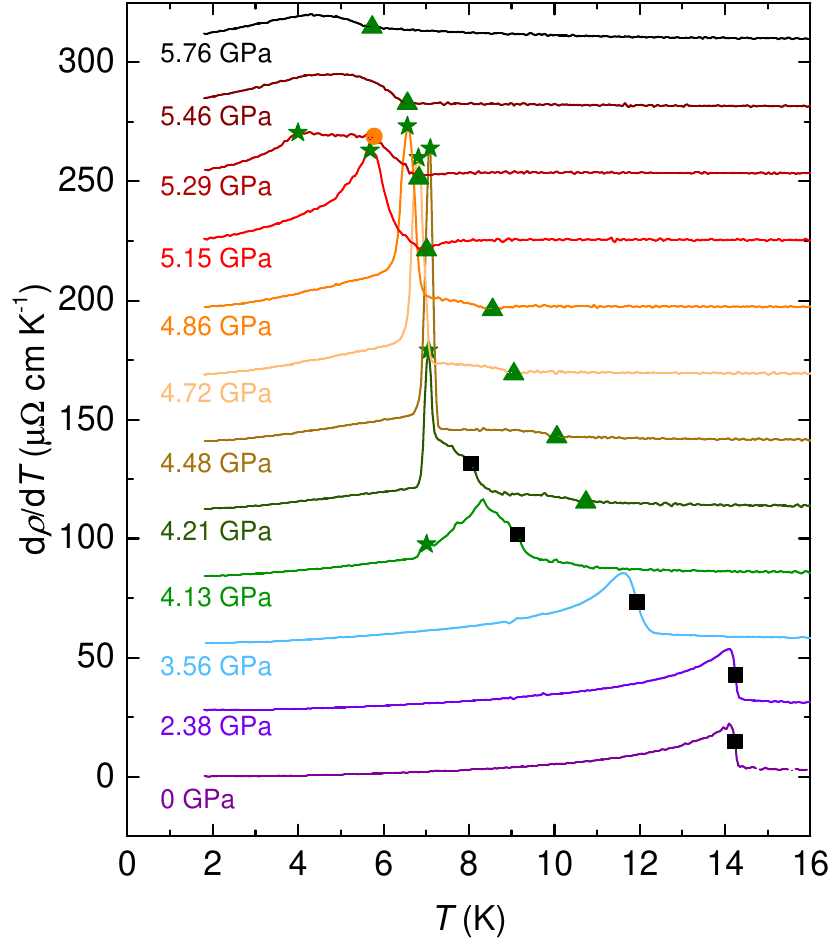}
			\end{center}
			\caption{\label{drho}(Color online) Evolution of the temperature derivative of the resistivity at low temperature for representative pressures. The data are vertically offset by 28\,$\mu\Omega$\,cm\,K$^{-1}$ to reduce overlap. Solid symbols represent the criteria described in Fig.\ref{Rho_T20K}. At 5.29\,GPa there is an additional anomaly in the d$\rho$/d$T$ as shown by the orange circle. } 
		\end{figure}

		Figure\,\ref{Rho_T20K} shows the evolution of the low temperature resistivity and its temperature derivatives in three selected pressure regions; (I) $p$\,$<$\,4.1\,GPa (II) 4.1\,GPa\,$<$\,$p$\,$<$\,5.3\,GPa and (III) $p$\,$>$\,5.3\,GPa. Below 4.1\,GPa the FM transition is seen as a sharp change of slope in the resistivity and transition temperature is obtained from the sharpest increase of d$\rho$/d$T$ (black square) (Figs.\,\ref{Rho_T20K}\,(a)-(b)). The FM transition temperature initially shows a weak increase with pressure and then decreases with further applied pressure up to 4.1\,GPa. Between 4.1-5.3\,GPa, the onset of magnetic transition 1 (\MPa) and magnetic transition 2 (\MPb) are revealed as a kink/upturn and a sharp drop in the $\rho(T)$ as shown in Figs.\,\ref{Rho_T20K}\,(c)-(e) . This can be clearly seen in the temperature derivative of the resistivity. Transition temperatures of PM-\MPa{} and \MPa{}-\MPb{} are obtained from the kink/minimum (green up-triangle, Figs.\,\ref{Rho_T20K}\,(c)-(e))  and sharp peak (green star) in d$\rho$/d$T$ (Figs.\,\ref{Rho_T20K}\,(c)-(e)) respectively. Although the magnetic ordering wave vector of \MPa{} is unknown, the  feature in the resistivity is similar to that associated with superzone gap formation\,\cite{Mackintosh1962PRL} and suggests an AFM nature for \MPa.  Both \MPa{} and \MPb{} transitions are observed between 4.1 to 5.3\,GPa and thermal hysteresis in $\rho$ for \MPb{} up to 5.3\,GPa (inset of Figs.\,\ref{Rho_T20K}\,(c)-(e)) indicates a first-order nature for this transition. On further increase of pressure, above 5.3\,GPa, \MPb{} disappears and a new magnetic transition, \MPc{}, continue to decrease with the increase of pressure  and  no thermal hysteresis is observed (Figs.\,\ref{Rho_T20K}\,(f)-(g)). Although features in the $\rho$(T) corresponding to the \MPa{} and \MPc{} transitions look similar, it is unclear whether it is same phase or not. Figure\,\ref{drho} shows the evolution of the temperature derivative of the resistivity for representative pressures. Solid symbols represent the criteria described in Fig.\,\ref{Rho_T20K}.

			\begin{figure}[htb!]
				\begin{center}
					\includegraphics[width=8.5cm]{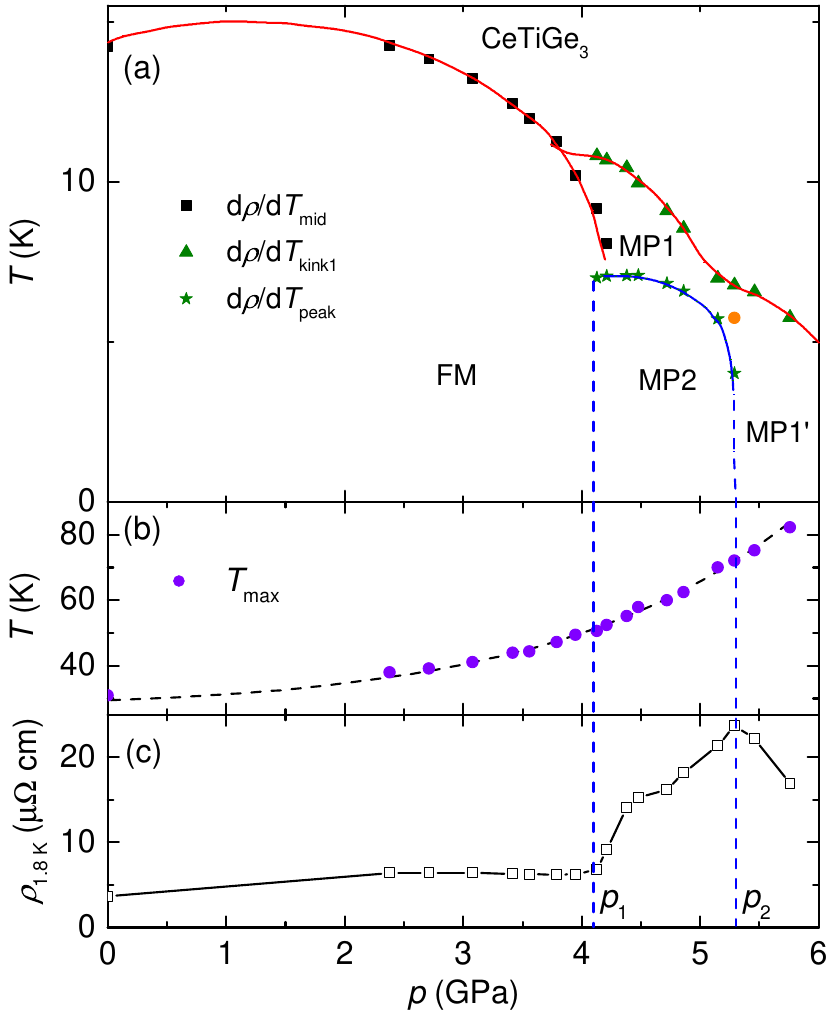}
				\end{center}
				\caption{\label{Phase_1}(Color online)(a) $T-p$ phase diagram of \CeTiGe{} in zero applied field. Transition temperatures are determined from the anomalies in d$\rho$/d$T$ as shown in Fig.\,\ref{Rho_T20K} and Fig.\,\ref{drho} . The values of critical pressure $p_1$ and $p_2$ are 4.1 and 5.3\,GPa respectively. Solid lines are guide to the eye and dashed lines are suggested extrapolations of phase boundaries. The red and blue color line represent the second and first order phase transitions. (b) Maximum in resistivity, $T_\text{max}$ (shown in Fig.\,\ref{Rho_T}(a)), as a function of pressure. (c) Pressure dependence of the $\rho$ at 1.8\,K.} 
			\end{figure}

		The temperature-pressure ($T-p$) phase diagram of \CeTiGe{} obtained from the resistivity measurements, is summarized in Fig.\,\ref{Phase_1}\,(a). At low pressures, the Curie temperature of the ambient pressure, FM phase (solid squares) shows a very weak pressure dependence and then decreases with pressure. For 4.1\,GPa\,$\leq$\,$p$\,$\leq$\,5.3\,GPa, there is an evidence for two phase transitions, \MPa{} and \MPb{} in the $\rho(p,T)$ curves, which interrupted the initial FM phase transition line. A similarly complex $T-p$ phase diagram has been observed in CeNiSb$_3$\,\cite{Sidorov2005PRB} and the recently studied itinerant ferromagnet LaCrGe$_3$\,\cite{Taufour2016PRL}. Pressure induced transitions from FM to AFM state are also observed in several other Ce-based compounds, such as CeAgSb$_2$\,\cite{Sidorov2003PRB},  CeNiSb$_3$\,\cite{Sidorov2005PRB}, CePd$_2$Ge$_3$\,\cite{Burghardt1997}, Ce$_2$Ni$_5$C$_3$\,\cite{Yamada2010} and CeRuPO\,\cite{Kotegawa2013JPSJ}. Above 5.3\,GPa, the low temperature \MPb{} phase disappears and \MPc{} continue to decrease with the increase of pressure. As mentioned above, it is unclear whether there is a phase boundary between \MPa{} and \MPc{} near 5.3\,GPa.

		In addition to the $T-p$ phase diagram, we find that, $T_\text{max}$ monotonically increases from  31\,K to 82\,K upon increasing pressure (Fig.\,\ref{Phase_1}\,(b)). The smooth change of $T_\text{max}$ indicates that the existence of the new phases is not associated with a discontinuous changes in the electronic or crystal structure or CEF splitting. Figure\,\ref{Phase_1}\,(c) shows the pressure evolution of the resistivity at 1.8\,K. The results show breaks in $\rho_{1.8 K}(p)$ at $p_1$ (FM to \MPb)  and a maximum at $p_2$ (\MPb{} to \MPa). The exact nature of the phase transitions at $p_1$ and $p_2$ are not known and to resolve this, it would be useful to study the magnetic ordering wave vector under pressure.

			\begin{figure}[htb!]
				\begin{center}
					\includegraphics[width=8.5cm]{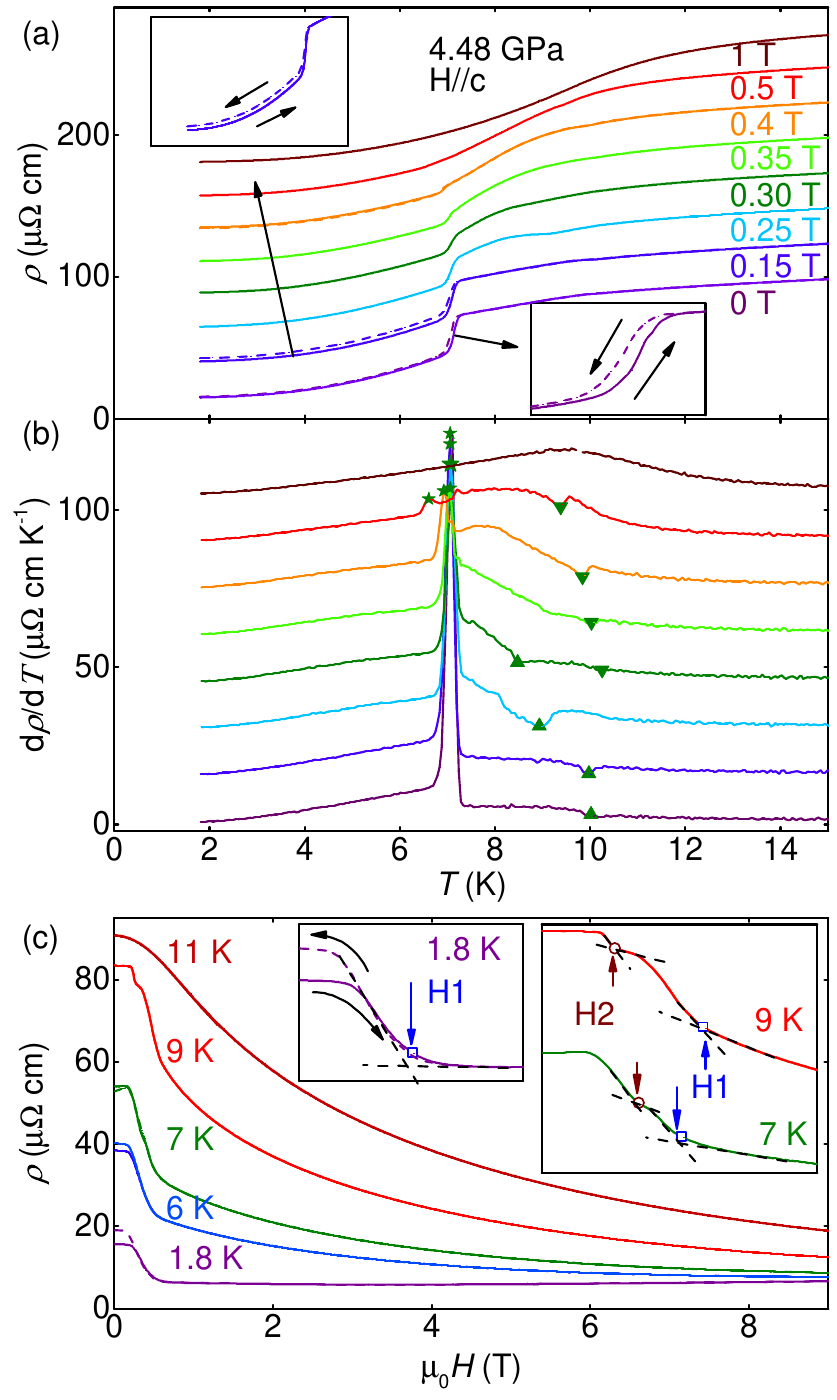}
				\end{center}
				\caption{\label{448Final}(Color online)(a) Temperature dependence of the resistivity at various fixed fields for $p$\,=\,4.48\,GPa and $H$\,$||$\,$c$. The data are vertically shifted by integer of  25\,$\mu\Omega$\,cm to avoid overlapping. The insets show the observed hysteretic behavior in temperature scan. Continuous and dashed lines represent the temperature increasing and decreasing respectively. (b) Corresponding temperature derivative (d$\rho$/d$T$) of (a). The data are vertically shifted by integer of  15\,$\mu\Omega$\,cm\,K$^{-1}$ to avoid overlapping. Solid symbols represent the criteria use to obtain the teansition temperatures at various magnetic fields. (c) Field dependence of  the resistivity  at fixed temperatures. For these data the sample was cooled in zero field and then $\rho$($H$) data was collected for increasing field ($\rho_\textrm{up}$) and then decreasing field ($\rho_\textrm{down}$). Then increase the temperature to the desired value and data was collected for increasing and decreasing field. Continuous and dashed lines represent the field increasing and decreasing respectively. Insets show the observed hysteretic behavior and the criteria used to obtain the transition fields. Above 7\,K no hysteretic behavior is observed.} 
			\end{figure}

		Application of an external magnetic field adds another dimension to our phase diagram and different behavior of the resistivity  anomalies under magnetic field allow us to explore further new phase regions of this material. Figure\,\ref{448Final}\,(a) shows the temperature dependence of $\rho$ at different magnetic fields, applied along the $c$-axis, at 4.48\,GPa. The sharp drop in the resistivity at low fields ($\mu_0H$\,$\leq$\,0.3\,T) broadens at higher fields. These data manifest hysteretic behavior up to 0.5\,T, indicating the first order nature of the transition. The zero-field kink in the resistivity, at 9.8\,K, changes into a hump with the increase of field (0.25\,T) and disappears at 0.3\,T. Another hump like feature appears above 0.35\,T and broadens with further increase of the field.  These features can be clearly observed in temperature derivative shown in Fig.\,\ref{448Final}\,(b). 
		
		The field dependence of $\rho$\,($p$\,=\,4.48\,GPa) below 7\,K shows a metamagnetic transition with a low field plateau followed by a step-like feature and  develops into two transitions above 7\,K (Fig.\,\ref{448Final}\,(c)). The solid and dashed lines represent the field increasing ($\rho_\text{up}(H)$) and decreasing ($\rho_\text{down}(H)$) respectively. The difference between $\rho_\text{up}(H)$-$\rho_\text{down}(H)$ shows a sizable deviation ($\rho$ is smaller in the increasing-field than the decreasing-field) for  0\,$\leq$\,$H$\,$\leq$\,0.3\,T range. In Fig.\,\ref{448Final}, hysteresis is apparent not only in the transition temperature (Fig.\,\ref{448Final}\,(a)) and transition field (Fig.\,\ref{448Final}\,(c)), but also in the magnitude of the resistivity. Similar hysteretic behavior is observed in the  CeAuSb$_2$\,\cite{Balicas2005PRB,Lorenzer2013,Zhao2016PRB} and Ce$T$Al$_4$Si$_2$ ($T$\,=\,Rh, Ir)\,\cite{Maurya2016JPSJ} . Based on the hysteretic behavior, we can conclude these metamagnetic transitions are likely associated with a first-order phase transition. The observed hysteresis in the magnitude of resistivity indicates the possibility of magnetic domains.   At temperatures above 11\,K, the resistivity shows a very broad anomaly and no transition has been observed. Criteria used to obtain transition fields are shown in the inset of Fig.\,\ref{448Final}\,(c). 
		
		\begin{figure}[htb!]
			\begin{center}
				\includegraphics[width=8.5cm]{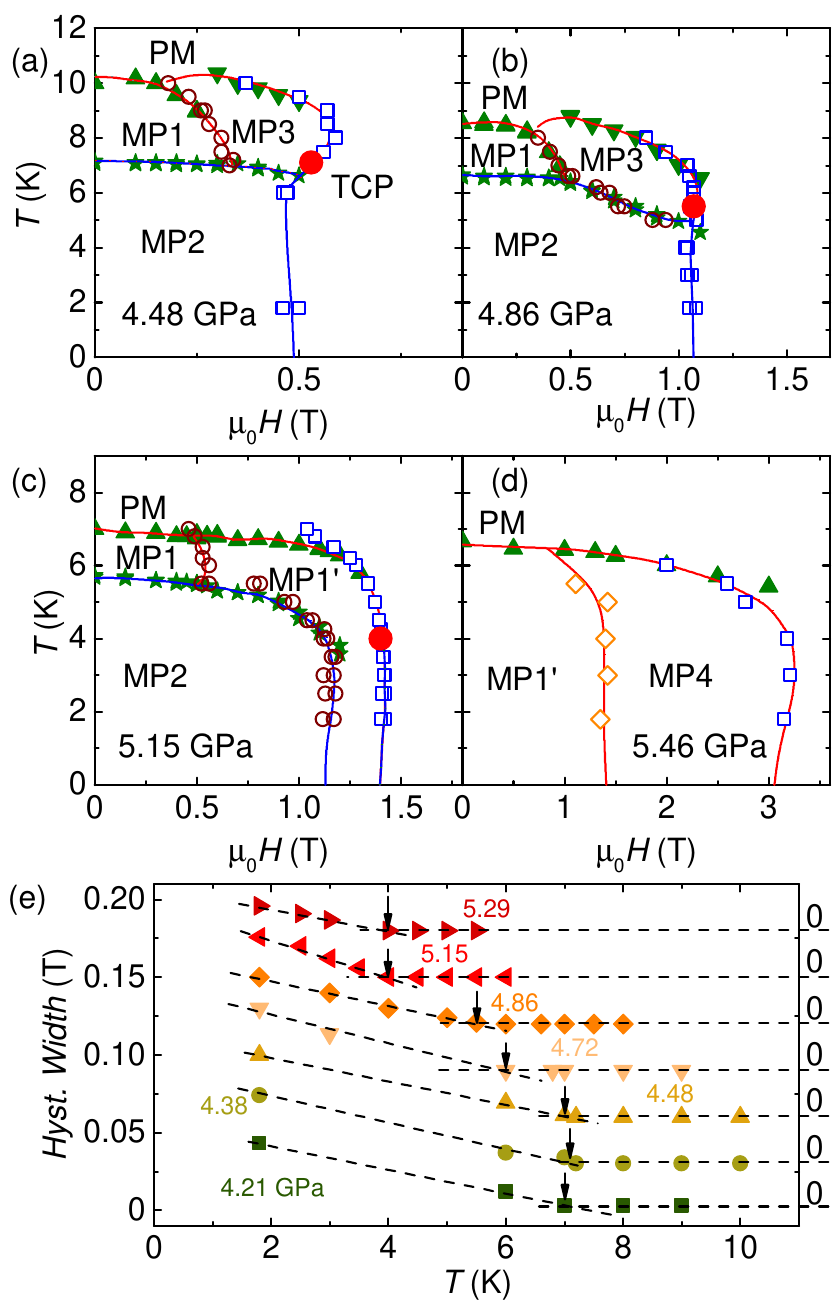}
			\end{center}
			\caption{\label{HT_comp}(Color online) $T-H$ phase diagrams for various pressures: (a) 4.48\,GPa (b) 4.86\,GPa (c) 5.15\,GPa and (d) 5.46\,GPa, determine by tracking various anomalies in temperature and field derivatives of resistivity measurement as shown by Fig.\,\ref{448Final}. Solid and open symbols represent transition temperatures determined by $T$-sweeps and transition fields determined by $H$-sweeps (as described in Fig.\,\ref{448Final}) respectively. Continuous blue and red lines indicate the first order and second order transitions respectively. See appendix for $T-H$ phase diagrams for all the measured pressures. (e) Temperature dependence of hysteresis widths for the transition at H1 at various pressures. The data are vertically offset by 0.03\,T to avoid overlap. Vertical arrows represent the estimated tricritical points for each pressure. Zero for each data set shown on right-hand axis.} 
		\end{figure}

		Figures\,\ref{HT_comp}\,(a)-(d) show the $T-H$ phase diagrams at representative pressures. Transition temperatures determined by $T$-sweep measurements are shown by closed symbols and anomalies appeared in isothermal $H$-sweep measurements are shown by open symbols. Continuous blue and red lines indicate the first order and second order transitions respectively (based on the presence or lack of hysteretic behavior respectively). The red circle represents the tricritical point (TCP) determined by Fig.\,\ref{HT_comp}\,(e). Temperature dependence hysteresis widths for the transition at H1 are shown in Fig.\,\ref{HT_comp}\,(e). The data are vertically offset by 0.03\,T to avoid overlap. Clear hysteresis at low temperature gradually decreases with increasing temperature and disappears at a TCP as shown by a vertical arrow.  In contrast to the wing-critical-point (WCP) in UGe$_2$\,\cite{Taufour2010PRL} and LaCrGe$_3$\,\cite{Kaluarachchi2017NatComm}, here we observed a TCP in the $T-H$ phase diagram where first order transition changes into the second order transition. This TCP corresponds to the boundary of the wing structure similar to UGe$_2$\,\cite{Taufour2010PRL} and LaCrGe$_3$\,\cite{Kaluarachchi2017NatComm}. The $T-H$  phase diagrams of \CeTiGe{} for pressures between 4.1\,-\,5.3\,GPa show complex behavior. Three magnetic phases (\MPa{}, \MPc{} and \MPb{}) are identified by the anomalies in the resistivity measurement. Both \MPa{} and \MPc{} phases are separated by \MPb{} phase by a first order transition as shown in Figs.\,\ref{HT_comp}\,(a)-(c). For pressures between 4.1-5.3\,GPa, these $T-H$  phase diagrams are similar to  those found for CeRu$_2$Al$_2$B\,\cite{Baumbach2012PRB}, which undergoes a second order AFM transition that is  followed by a first order FM transition as a function of temperature. Above 5.3\,GPa, only two magnetic phases; \MPc{} and MP4 are identified by the resistivity measurements and there is no longer a first order phase transition boundary observed. The $T-H$ phase diagrams for all the pressures above 4.13\,GPa are shown in Fig.\,\ref{T_H_all}.

		\begin{figure}[htbp!]  
			\begin{center}
				\includegraphics[width=8.5cm]{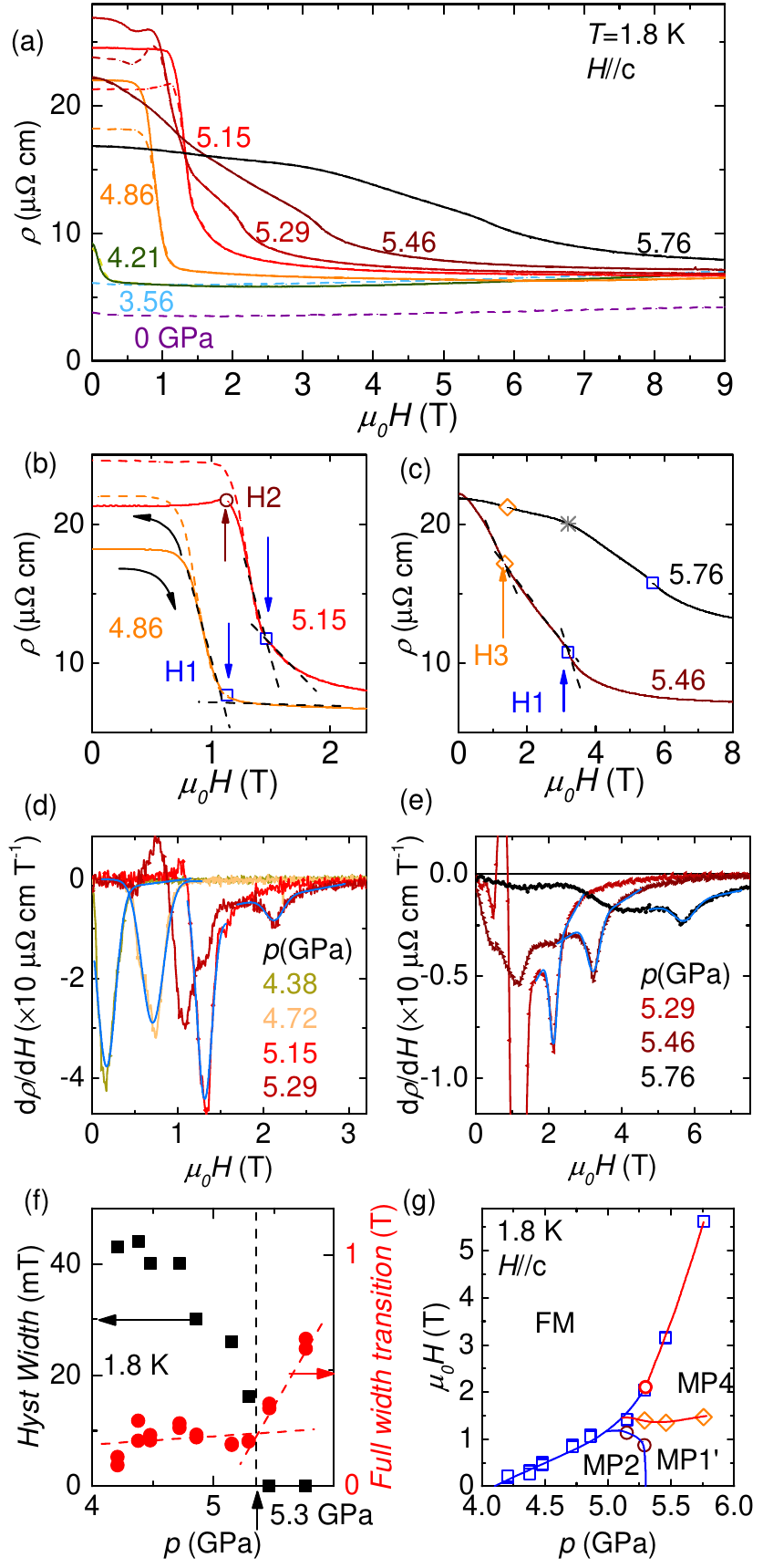}
			\end{center}
			\caption{\label{RH}(Color online) (a) Field dependence of $\rho$ at 1.8\,K for various pressures. Continuous and dashed lines represent the field increasing and decreasing respectively. Representative $\rho(H)$ data for (b) 4.1\,GPa\,$<$\,$p$\,$<$\,5.3\,GPa, (c) $p$\,$>$\,5.3\,GPa and the criteria used to obtain the transition fields at 1.8\,K. The open symbols represent the corresponding transition fields. The gray star represents the shoulder like anomaly appeared at 5.76\,GPa (see Appendix for more details). (d)-(e) Representative derivative, d$\rho$/d$H$ data for H1 transition at 1.8\,K. The blue color lines represent the Gaussian+linear-background fitted curves which are used to obtained full-width of the H1 transition. (f) left axis shows the pressure dependence hysteresis width of transition H1 at 1.8\,K. Right axis shows pressure dependence of the full-width of H1 obtained by d$\rho$/d$H$ ((d)-(e)) at 1.8\,K. Vertical dashed line represent the tricritical pressure\,$\sim$\,5.3\,GPa, at 1.8\,K. (g) $H-p$ phase diagram at 1.8\,K based on the criterion shown in (b,c). Blue and red solid lines represent the first and second order transitions. Red open circle represents the extrapolated QTCP.} 
		\end{figure}

		Figure\,\ref{RH}\,(a) shows the field dependence of the resistivity at 1.8\,K, $\rho(H)$, for  different pressures. For the pressures in between $p_1$ and $p_2$, $\rho(H)$ for an increasing magnetic field shows a clear metamagnetic transition with a substantial ($>$\,$40\%$), drop of resistivity. For higher pressures, the sharp drop in the  $\rho(H)$ disappears and several metamagnetic transitions can be observed. Figures\,\ref{RH}\,(b) and (c) show the representative magnetoresistance data for 4.1\,GPa\,$<$\,$p$\,$<$\,5.3\,GPa and $p$\,$>$\,5.3\,GPa respectively. Transition fields determined by $H$-sweeps measurements are shown by the open symbols. To estimate the transition width, we used the  field derivative of the resistivity at 1.8\,K, as shown in Figs.\,\ref{RH}(d) and (e). The minimum at H1 is fitted with Gaussian+linear-background and obtained the width of the Gaussian distribution. The blue color lines in Figs.\,\ref{RH}(b) and (c) represent the fitted curves to the data. We noticed that the  transition width (Fig.\,\ref{RH}\,(f) right axis) at H1 at 1.8\,K remains small for the first-order transition and becomes broad in the second-order regime. Using linear extrapolation as represented by red dashed lines, we obtained pressure corresponding to the TCP at 1.8\,K, which is 5.3\,GPa. In addition to that, the temperature dependence hysteresis width for transition H1 at 1.8\,K is also suppressed with the pressure and disappeared above 5.3\,GPa as shown in Fig.\,\ref{RH}\,(f) left axis. Figure\,\ref{RH}\,(g) shows the $H-p$ phase diagram at 1.8\,K constructed from the above criteria. The magnetic field that corresponds  to the H1 transition is shifted up with pressure. Its extrapolation down to zero yields $p$\,$\cong$\,4.1\,GPa, which is in agreement with the $p_1$ obtained from $T$-$p$ diagram (Fig.\,\ref{Phase_1}\,(a)). We observe the increasing rate of metamagnetic transition field with respect to pressure, changes near 5.3\,GPa. Similar $H-p$ phase diagrams at low temperature have been observed in LaCrGe$_3$\,\cite{Kaluarachchi2017NatComm} and CeRu$_2$(Si$_{1-x}$Ge$_x$)$_2$ system\,\cite{Matsumoto2011,Aoki2014JPSJ}. CeRu$_2$Ge$_2$ is a local moment system\,\cite{Sullow1999PRL}, while CeRu$_2$Si$_2$ is itinerant\,\cite{Aoki1993PRL}. Application of pressure to  CeRu$_2$Ge$_2$ gives nearly same magnetic phase diagram as that of  CeRu$_2$(Si$_{1-x}$Ge$_x$)$_2$\,\cite{Wilhelm1998,Haen1999}. Observed transport and de Haas-van Alphen data suggest that, for this system, change of the $f$\,-\,electron nature from local to itinerant occurs when the FM phase disappears\,\cite{Matsumoto2011}. On the other hand, itinerant ferromagnet LaCrGe$_3$ show tricritical wings as well as modulated magnetic phase. Interestingly, $T-p-H$ phase diagram of both LaCrGe$_3$\,\cite{Kaluarachchi2017NatComm} and CeRu$_2$Ge$_2$\,\cite{Aoki2014JPSJ} without AFM states is similar to the itinerant weak ferromagnet like UGe$_2$\,\cite{Taufour2010PRL}. This similarity might imply that the physics behind these phase diagrams are not very different.

\begin{figure}[htb!]
			\begin{center}
				\includegraphics[width=8.5cm]{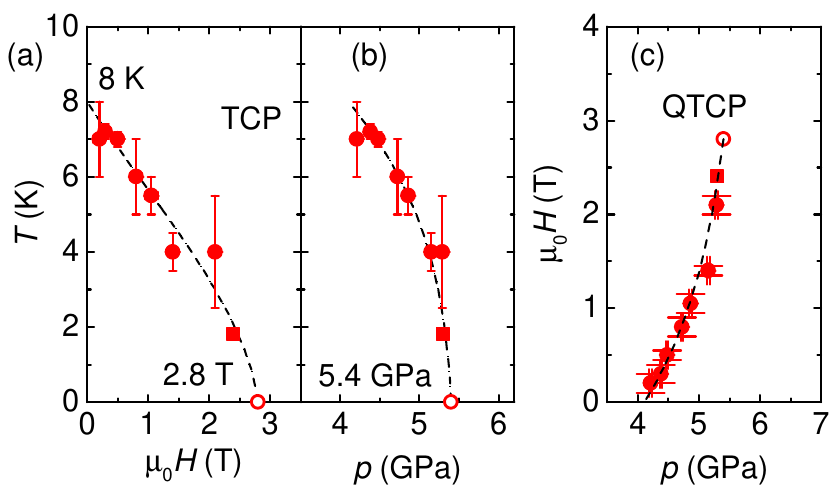}
			\end{center}
			\caption{\label{Hyst_width_3}(Color online) Projection of wings in (a) $T-H$  (b) $T-p$ and (c) $H-p$ planes. Red solid circles represent the TCP determined by  Fig.\,\ref{HT_comp}(e). Red solid squared obtained from Fig.\,\ref{RH}\,(f). Dashed lines are guides to the eyes and open red circles represent the extrapolated QTCP.} 
		\end{figure}

		The projection of the wing lines in $T-H$, $T-p$ and  $H-p$  planes are shown in Figs.\,\ref{Hyst_width_3}\,(a),(b) and (c) respectively. The wing lines can be extrapolated to a quantum-tri-critical-point (QTCP) at 0\,K, which is found to be at 2.8\,T at 5.4\,GPa. Theoretical analysis based on Landau expansion shows that the slope of the wings d$T$/d$H$ and d$p$/d$H$ are infinite near $H$\,=\,0\,T\,\cite{Taufour2016PRB}. This was observed experimentally in URhGe\,\cite{Nakamura2017PRB}. It was also observed in LaCrGe$_3$, despite the existence of another magnetic phase\,\cite{Kaluarachchi2017NatComm}. Here, we do not observe such behavior which could be due to the existence of the magnetic phase \MPa{} or to the lack of data near $p_1$. More careful measurements near $p_1$.  are required. Also, the TCP at $H$\,=\,0\,T is found to be $\sim$\,8\,K and this is below the \MPa{} transition. A similar observation was made in LaCrGe$_3$\,\cite{Kaluarachchi2017NatComm} where the TCP seems to be located below the Lifshitz point. Recent theoretical description by Belitz and Kirkpatrick in Ref.\,\onlinecite{Belitz2017arXiv} shows the complex behavior of the phase diagrams of metallic magnets when an AFM order is observed in addition to the FM phase due to the quantum fluctuations. Similar to the Fig.\,4\,(a) in Ref.\,\onlinecite{Belitz2017arXiv}, we observed a QTCP where first order AFM-FM transition changes into the second order AFM-FM transition at 2.8\,T at 5.4\,GPa (see Fig.\,\ref{RH}\,(g)). Very recently QTCP has experimentally observed in NbFe$_2$\,\cite{Friedemann2017NatPhy}.

		\begin{figure}[htb!]
			\begin{center}
				
				\includegraphics[width=8.5cm]{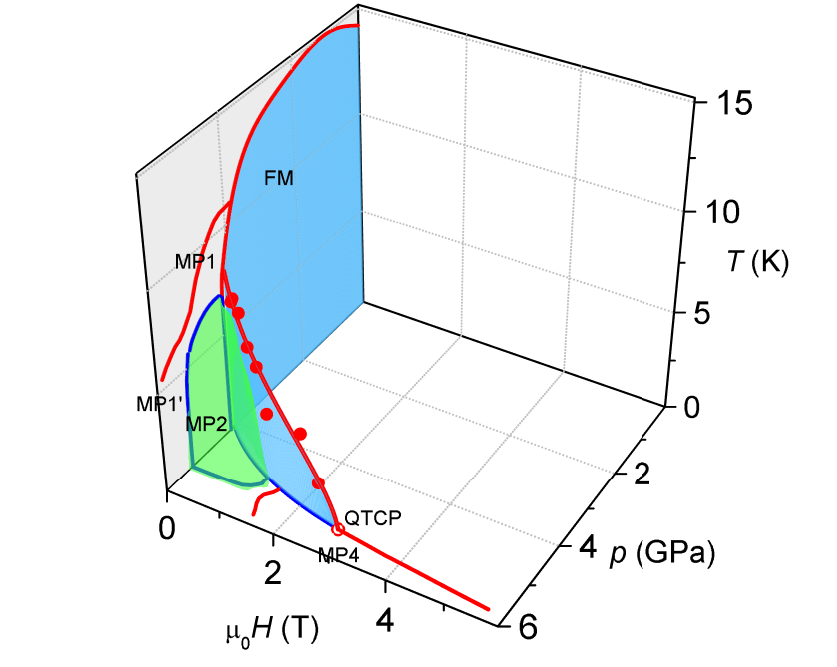}
			\end{center}
			\caption{\label{3D}(Color online) The constructed, partial, $T-p-H$ phase diagram of CeTiGe$_3$ based on resistivity measurements. Blue color surfaces represent the first-order planes and green color surface represents the first-order \MPb{} phase boundary. Continuous red and blue lines represent the second and first order transition respectively. The open circle represent extrapolated QTCP.} 
		\end{figure}		
		
		The constructed, partial,  $T-p-H$ phase diagram of CeTiGe$_3$ based on resistivity measurements is shown in Fig.\,\ref{3D}. A FM QCP in CeTiGe$_3$ is avoided by the appearance of \MPa{} and \MPb{} phases, and shows field induced wing structure above 4.1\,GPa.  The estimated QTCP is shown by the open red circle.  In order to provide clear picture of the wing structure phase diagram, we show only selected phases here (see Fig.\,\ref{T_H_all} for $H$-$T$ phase diagrams at various pressures). In the case of the itinerant the ferromagnet, LaCrGe$_3$\,\cite{Taufour2016PRL,Kaluarachchi2017NatComm}, the second-order FM transition becomes a first order at a tricritical point in the $T$-$p$ plane and application of a magnetic field reveals a wing structure phase diagram. Appearances of modulated magnetic phase in LaCrGe$_3$\,\cite{Kaluarachchi2017NatComm} makes it the first example of new type of phase diagram of metallic quantum ferromagnets. Unlike LaCrGe$_3$ (Fig.\,5 in Ref.\,\cite{Kaluarachchi2017NatComm}), where, wings are extended beyond the AFM phases,  the observed wings in \CeTiGe{} are always bounded by the AFM phases. This can be clearly visualized in Fig.\,\ref{RH}\,(g) (for comparison see Fig.\,4 in Ref.\,\cite{Kaluarachchi2017NatComm}). The observation of QTCP in metallic magnets in the case of appearance of AFM order in addition to the FM order is theoretically described by Belitz and Kirkpatrick\,\cite{Belitz2017arXiv}. This theoretical finding is consistent with our experimental observation of QTCP in \CeTiGe{}. Therefore, CeTiGe$_3$ is a good example of a Ce-based compounds in which the system can be driven into various magnetic ground state by fine tuning of the exchange interaction achieved by temperature, pressure and magnetic field.

		\section{Conclusions}
		We have measured the high pressure electrical resistivity of \CeTiGe{} up to 5.8\,GPa and 9\,T and found a complex $T-p-H$ phase diagram. The ferromagnetic transition at ambient pressure initially slightly increases and then decreases, indicates that CeTiGe$_3$ is located just below the maximum (left side) of the Doniach phase diagram. The ferromagnetic transition suppresses near 4.1\,GPa and cascade of phase transitions are observed above that.  Change in residual resistivity near 4.1\,GPa and 5.3\,GPa suggests a modification of the electronic structure upon entering these magnetic phases. Thus,  \CeTiGe{}  is another clear example of avoided ferromagnetic quantum critical point due to appearance of magnetic phase (probably antiferromagnetic). Application of magnetic field under pressure above 4.1\,GPa reveals wing structure phase diagram. In contrast to the wing critical point in LaCrGe$_3$, we observed a tricritical point in $H$-$p$ plane, which corresponding to the boundary of the wing structure. Estimated quantum tricritical point of \CeTiGe{} is located at 2.8\,T at 5.4\,GPa. We believe that the present work will stimulate further experiments to investigate the properties of this material.

		\section*{ACKNOWLEDGMENTS}
		We would like to thank S.~Manni and A.~Kreyssig for useful discussions. This work was supported by the U.S. Department of Energy (DOE), Office of Science, Basic Energy Sciences, Materials Science and Engineering Division. The research was performed at the Ames Laboratory, which is operated for the U.S. DOE by Iowa State University under contract No. DE-AC02-07CH11358. V.T. was partially supported by Critical Material Institute, an Energy Innovation Hub funded by U.S. DOE, Office of Energy Efficiency and Renewal Energy, Advanced Manufacturing Office.	
	
	\appendix
	\label{Appendix}
	\section{Appendix}
	
\begin{figure*}[htb!]
		\begin{center}
			\includegraphics[width=16cm]{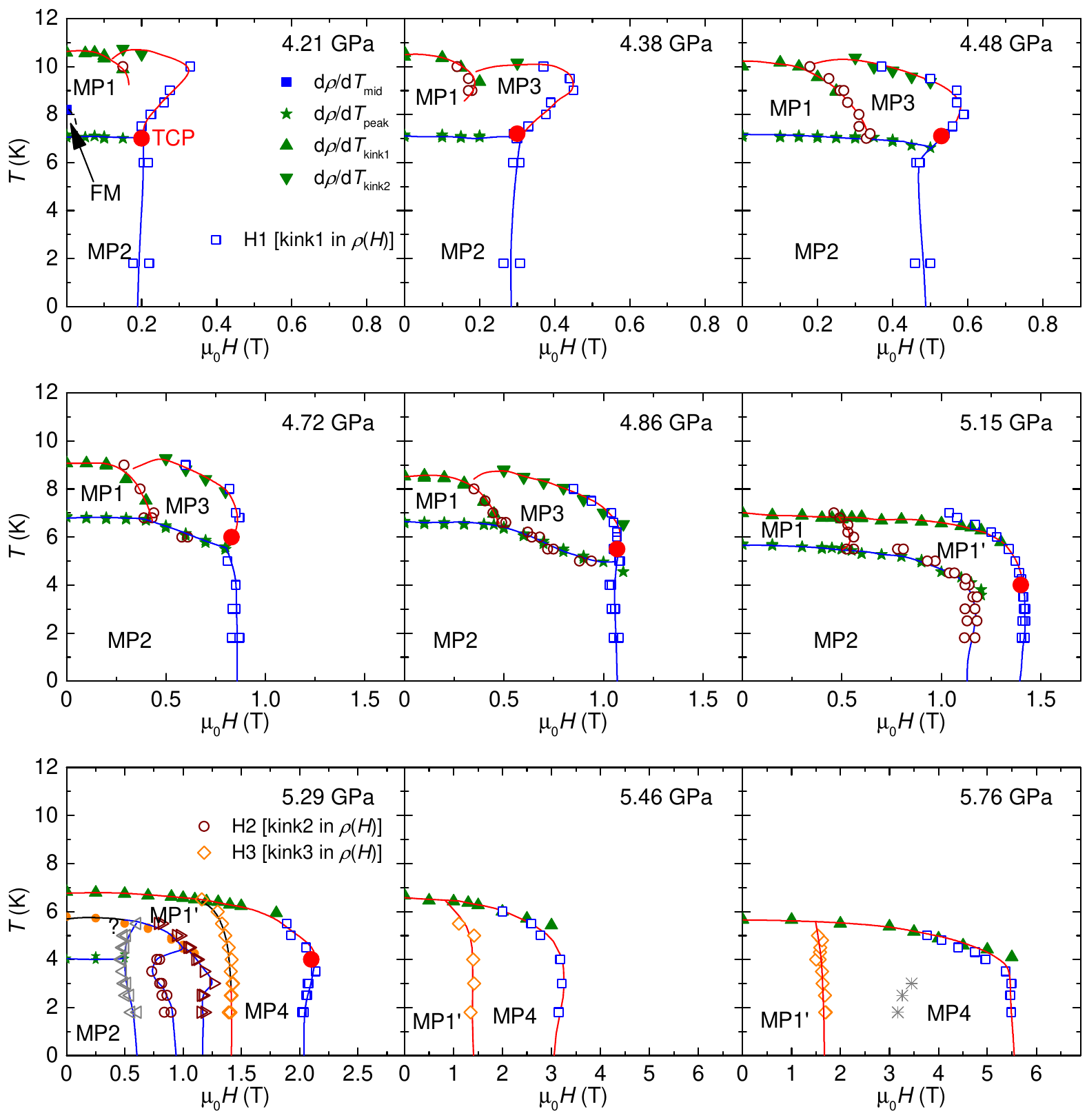}
		\end{center}
		\caption{\label{T_H_all}(Color online)  $T-H$ phase diagrams, including those shown in Fig.\,\ref{HT_comp}\,(a)-(d,) at various increasing applied pressures. At 5.29\,GPa, $T-H$ phase diagrams show a complex behavior and with additional metamagnetic transitions (gray and brown open triangles) in $\rho$($H$) data (raw data are not shown). H1, H2 and H3 are the anomalies observed in $\rho$($H$) data as shown in Fig.\,\ref{448Final}\,(c) and Figs.\,\ref{RH}\,(b)-(c).} 
	\end{figure*}

	Figure\,\ref{T_H_all} shows the constructed $T-H$ phase diagrams for pressures between 4.21 to 5.76\,GPa. There is a clear difference in the $T-H$ phase diagrams below 4.86\,GPa and above 5.46\,GPa.  $T-H$ phase diagram for the intermediate pressure, 5.29\,GPa, shows a complex behavior. Also, we observed an additional shoulder-like anomaly in $\rho$($H$) at 5.76\,GPa (gray color star in Fig.\,\ref{RH}\,(c) and Fig.\,\ref{T_H_all}).  When the temperature was increased, it became broadened and merged with H1 and no loner resolvable. H1, H2 and H3 are the anomalies observed in $\rho$($H$) data as shown in Fig.\,\ref{448Final}\,(c) and Figs.\,\ref{RH}\,(b)-(c)

\newpage	
\pagebreak

\bibliographystyle{apsrev4-1}
$*$ Current affiliation: Department of Physics, University of California, Davis, California 95616, USA.

%

\end{document}